\documentclass[letterpaper,twocolumn,reprint,amsmath,amssymb,prl]{revtex4-2}
\usepackage{mathptmx}

\usepackage[T1]{fontenc}
\usepackage[latin9]{inputenc}
\usepackage{color}
\usepackage{amsmath}
\usepackage{graphicx}
\usepackage[pdfusetitle,
 bookmarks=true,bookmarksnumbered=false,bookmarksopen=false,
 breaklinks=false,pdfborder={0 0 0},pdfborderstyle={},backref=false,colorlinks=true]
 {hyperref}

\makeatletter

\pdfpageheight\paperheight
\pdfpagewidth\paperwidth

\makeatother

\begin{document}
\title{Controlling Excitation Localization in Waveguide QED Systems}
\author{C.-Y. Lee$^{1}$}
\email{lcy511085@gmail.com}
\author{K.-T. Lin$^{2}$}
\email{alex968.tw@gmail.com}
\author{G.-D. Lin$^{1,2,3}$}
\author{H. H. Jen$^{4,1}$}

\affiliation{$^{1}$Physics Division, National Center for Theoretical Sciences,
Taipei 10617, Taiwan~\linebreak{}
$^{2}$Trapped-Ion Quantum Computing Laboratory, Hon Hai Research
Institute, Taipei 11492, Taiwan~\linebreak{}
$^{3}$Department of Physics and Center for Quantum Science and Engineering,
National Taiwan University, Taipei 10617, Taiwan~\linebreak{}
$^{4}$Institute of Atomic and Molecular Sciences, Academia Sinica,
Taipei 10617, Taiwan}

\begin{abstract}
We theoretically investigate excitation dynamics in one-dimensional
arrays of quantum emitters coupled to a waveguide, focusing on localization
and long-time population trapping. By combining time-domain simulations
with spectral analysis of an effective non-Hermitian Hamiltonian,
we identify two distinct mechanisms that give rise to localization:
geometry-induced subradiance and disorder-induced Anderson-like confinement.
Spatially modulated emitter arrangements\textemdash such as single-
and double-Gaussian transverse profiles\textemdash enable long-lived
subradiant modes even in the absence of disorder, with decay rates
that can be finely controlled via geometric parameters. In contrast,
localization in uniform arrays emerges only when disorder breaks spatial
symmetry and suppresses collective emission through interference.
We track the crossover between geometric and disorder-induced regimes,
finding that double-Gaussian profiles exhibit clear spatial signatures
of this transition, while single-Gaussian configurations display more
gradual changes. These results establish geometry and disorder as
complementary tools for engineering long-lived quantum states in waveguide
QED systems, with direct relevance for scalable implementations in
photonic platforms.
\end{abstract}
\maketitle
\hypersetup{
colorlinks=true,
urlcolor=blue,
linkcolor=blue,
citecolor=blue
}

\section{Introduction}

Waveguide quantum electrodynamics (QED) provides a versatile platform
for exploring light\textendash matter interactions in low-dimensional,
open quantum systems \citep{LeKien2005,Loo2013,Arcari2014,Douglas2015,Goban2015,Pichler2015,Shahmoon2016,Ruostekoski2016,Lodahl2017,Chang2018,Mahmoodian2020,LeJeannic2021,Iversen2022,Sheremet2023,GonzalezTudela2024}.
Arrays of quantum emitters coupled to one-dimensional (1D) photonic
reservoirs exhibit rich collective phenomena, including superradiance,
subradiance, and photon-mediated long-range interactions \citep{Pichler2015,Chang2018,Dinc2019,Fedorovich2022,CardenasLopez2023,Joshi2023}.

A central challenge in such systems is understanding and controlling
excitation localization, i.e., the long-time retention of population
within the emitter array. In conventional models, localization typically
arises from disorder-induced interference, as in Anderson localization
\citep{Anderson1958}. However, in open systems with non-Hermitian
dynamics, engineered geometry can also induce localization via spatially
inhomogeneous coupling, leading to the formation of subradiant modes
\citep{Scully2015,Facchinetti2016,Jen2016,Sutherland2016,Plankensteiner2017,Jenkins2017,Jen2017,Bhatti2018,MorenoCardoner2019,Albrecht2019,Zhang2019,Ferioli2021,MorenoCardoner2022,Hsu2024,Jen2025}
that are protected from collective decay. Despite extensive studies
on disorder-induced localization in photonic and atomic systems \citep{Schwartz2007,Lahini2008,See2019,Jen2020,Jen2022,Sels2022,Wu2024},
the role of emitter geometry in waveguide QED remains underexplored.
Existing works often fix emitter positions, treating geometry as a
static design choice, rather than as a tunable parameter for engineering
localization.

In this work, we investigate population trapping in 1D arrays of quantum
emitters coupled to a waveguide, by performing a systematic comparison
of three spatial configurations: linear, single-Gaussian, and double-Gaussian
transverse profiles. Disorder is introduced along the propagation
axis, and localization behavior is analyzed through time-domain simulations
and spectral decomposition of an effective non-Hermitian Hamiltonian
\citep{Giusteri2015}. The imaginary parts of the eigenvalues determine
the decay rates of collective modes, enabling the identification of
dominant long-lived eigenstates responsible for excitation retention.

Our results show that localization originates from two qualitatively
distinct mechanisms: geometry-induced subradiance and disorder-induced
localization, with a crossover behavior observed as disorder strength
increases. The transition is subtle in single-Gaussian structures,
but becomes more pronounced in double-Gaussian configurations that
exhibit clear spatial signatures of the crossover. We further examine
how localization depends on system size and find that structured emitter
geometries\textemdash such as single- and double-Gaussian profiles\textemdash exhibit
greater robustness and sustained population trapping than uniform
linear arrays, especially in the weak-disorder regime.

These findings provide new insights into controllable radiative dynamics
in waveguide QED and offer practical design principles for suppressing
decoherence in quantum optical networks. The predicted behaviors are
accessible using current experimental platforms, including superconducting
qubits \citep{Roushan2017,Xu2018}, cold atoms coupled to nanophotonic
waveguides \citep{Kim2019,Luan2020}, and quantum dots in photonic
crystals \citep{Arcari2014}.

\section{Theoretical model}

\begin{figure}[t]
\begin{centering}
\includegraphics[width=8.8cm]{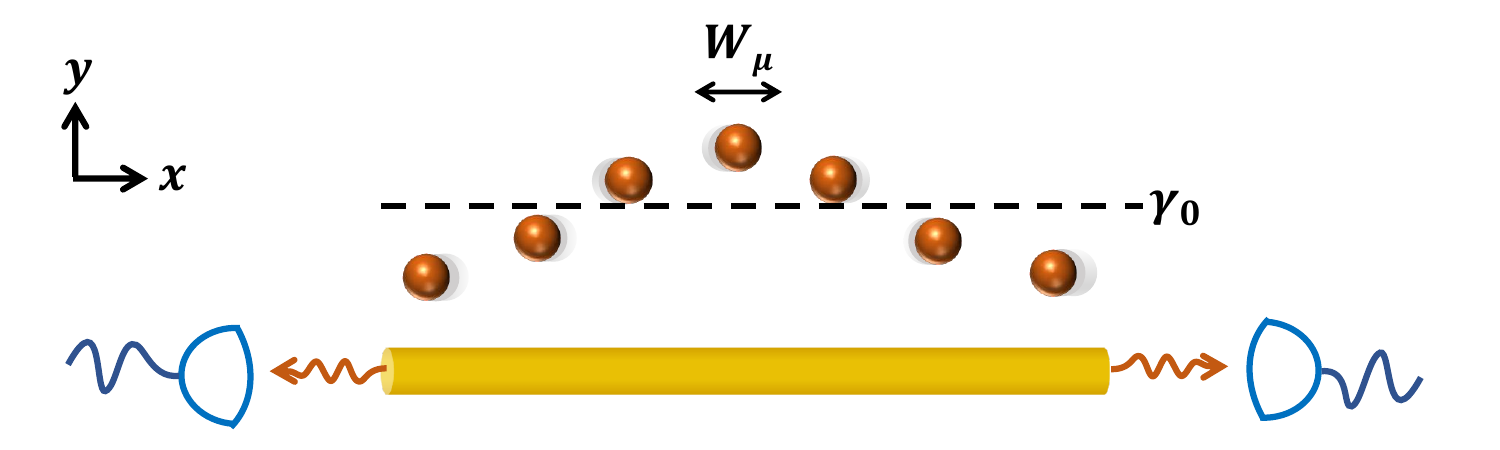}
\par\end{centering}
\caption{Schematic of the system under consideration: a one-dimensional array
of quantum emitters symmetrically coupled to a waveguide. Different
transverse emitter geometries are realized by introducing position-dependent
displacements $y_{\mu}.$ The coupling rate $\gamma_{0}$ is defined
with respect to a reference position relative to the waveguide, as
illustrated. \label{fig:model}}
\end{figure}

We consider a 1D array of $N$ identical two-level quantum emitters,
with ground state $\left|g\right\rangle $ and excited state $\left|e\right\rangle $,
separated by a transition frequency $\omega_{eg},$ and coupled to
a photonic waveguide, as illustrated in Fig.~\ref{fig:model}. In
this setup, the transverse positions $y_{\mu}$ of the emitters can
be varied to produce inhomogeneous waveguide-emitter couplings, allowing
control over the collective relaxation dynamics of the system. The
system's dynamics in the interaction picture are governed by the Lindblad
master equation \citep{GonzalezTudela2013,Jen2020a}:

\begin{align}
\frac{d\rho}{dt}= & -i\left[H,\rho\right]-\frac{\gamma_{0}}{4}\sum_{\mu,\nu=1}^{N}e^{-\left(y_{\mu}+y_{\nu}\right)/\xi_{c}}e^{-ik_{eg}\left|x_{\mu}-x_{\nu}\right|}\nonumber \\
 & \quad\times\left(\sigma_{\mu}^{\dagger}\sigma_{\nu}\rho+\rho\sigma_{\mu}^{\dagger}\sigma_{\nu}-2\sigma_{\nu}\rho\sigma_{\mu}^{\dagger}\right)\label{eq:master}
\end{align}
 with the coherent interactions described by 
\begin{equation}
H=-i\frac{\gamma_{0}}{4}\sum_{\mu\neq\nu}\sum_{\nu=1}^{N}e^{-\text{\ensuremath{\left(y_{\mu}+y_{\nu}\right)}}/\xi_{c}}\left(e^{ik_{eg}\left|x_{\mu}-x_{\nu}\right|}\sigma_{\mu}^{\dagger}\sigma_{\nu}-\text{H.c.}\right),\label{eq:Hamiltonian}
\end{equation}
 where $\sigma_{\mu}^{\dagger}\equiv\left|e\right\rangle _{\mu}\left\langle g\right|$
is the dipole raising operator, and $\gamma_{0}$ denotes the emitter-waveguide
coupling strength. This master equation is derived under the Born-Markov
approximation \citep{Lehmberg1970}, and applies to 1D photonic reservoirs
that mediate long-range dipole-dipole interactions \citep{Dicke1954}.
The parameters $\xi_{c}$ and $k_{eg}$ represent the characteristic
decay length and photon wavevector, respectively, where $k_{eg}\equiv\omega_{eg}/v$
and $v$ is the group velocity of the guided mode. For simplicity,
we set $\xi_{c}=1$ throughout this work, so that all lengths and
coupling expressions are normalized by the evanescent decay length.
The spatially inhomogeneous coupling arises from the evanescent nature
of the guided mode, which decays exponentially with transverse displacement
from the waveguide.

To characterize the interaction range, we define the dimensionless
parameter $\xi\equiv k_{eg}\left|x_{\mu+1}-x_{\mu}\right|,$ which
corresponds to the normalized inter-emitter spacing along the propagation
direction. The emitters are positioned uniformly along the propagation
axis, such that the inter-emitter spacing is fixed ($\xi=$ const).
However, their transverse positions $y_{\mu}$ are allowed to vary,
leading to position-dependent coupling strengths with the photonic
waveguide. This geometry preserves the one-dimensional topology of
the system, while enabling spatially inhomogeneous interactions through
transverse displacement.

We restrict our analysis to the single-excitation subspace, which
fully describes the system dynamics under initial single-excitation
conditions. The general state can be written as: $\left|\Psi\left(t\right)\right\rangle =\sum_{\mu=1}^{N}a_{\mu}\left(t\right)\left|\psi\right\rangle _{\mu},$
where $\left|\psi\right\rangle _{\mu}=\left|e\right\rangle _{\mu}\left|g\right\rangle ^{\otimes\left(N-1\right)}$
denotes a single excitation localized on the $\mu$-th emitter. Within
this subspace, the effective non-Hermitian Hamiltonian takes the form
\citep{Stannigel2012,Pichler2015}:
\begin{align}
H_{\text{eff}}= & -i\frac{\gamma_{0}}{4}\sum_{\mu\neq\nu}\sum_{\nu=1}^{N}e^{-\text{\ensuremath{\left(y_{\mu}+y_{\nu}\right)}}}e^{ik_{eg}\left|x_{\mu}-x_{\nu}\right|}\sigma_{\mu}^{\dagger}\sigma_{\nu}\nonumber \\
 & -i\frac{\gamma_{0}}{2}\sum_{\nu=1}^{N}e^{-2y_{\nu}}\sigma_{\nu}^{\dagger}\sigma_{\nu}.\label{eq:effective Hamiltonian}
\end{align}

To study the effect of imperfections, we introduce static disorder
$W_{\mu}\in\pi\left[-\bar{\omega},\bar{\omega}\right],$ modeling
onsite phase fluctuations along the waveguide axis. This leads to
a Schr\"odinger-type equation governing the time evolution of the excitation
amplitudes:
\begin{align}
\dot{a}_{\mu}\left(t\right)= & -\frac{\gamma_{0}}{2}\sum_{\mu<\nu}e^{-\text{\ensuremath{\left(y_{\mu}+y_{\nu}\right)}}}e^{i\left|\mu-\nu\right|\xi-i\left(W_{\mu}-W_{\nu}\right)}a_{\nu}\left(t\right)\nonumber \\
 & -\frac{\gamma_{0}}{2}\sum_{\mu>\nu}e^{-\text{\ensuremath{\left(y_{\mu}+y_{\nu}\right)}}}e^{i\left|\mu-\nu\right|\xi-i\left(W_{\nu}-W_{\mu}\right)}a_{\nu}\left(t\right)\nonumber \\
 & -\frac{\gamma_{0}}{2}e^{-2y_{\mu}}a_{\mu}\left(t\right).\label{eq:schrodinger equation}
\end{align}
In the following sections, we analyze the time evolution of $a_{\mu}\left(t\right)$
under various array configurations and examine how disorder influences
localization and excitation transport in geometrically structured
one-dimensional emitter arrays.

\section{geometry-induced localization and mode analysis}

In this section, we investigate how the geometry of emitter arrays
influences localization phenomena and long-time excitation trapping.
We consider a system of $N=101$ emitters with uniform inter-emitter
spacing $\xi,$ and prepare the initial state as a symmetric Dicke
state, in which the excitation is uniformly distributed across the
array.

\subsection{Localization Induced by Varying Inter-Emitter Spacing}

We first examine how localization depends on the inter-emitter spacing
$\xi,$ keeping the transverse displacement parameters fixed at $\eta=0.05$
and $\sigma=0.2$ for the single-Gaussian profile, and $\sigma=0.075$
for the double-Gaussian case.

\begin{figure}[h]
\begin{centering}
\includegraphics[width=8.8cm]{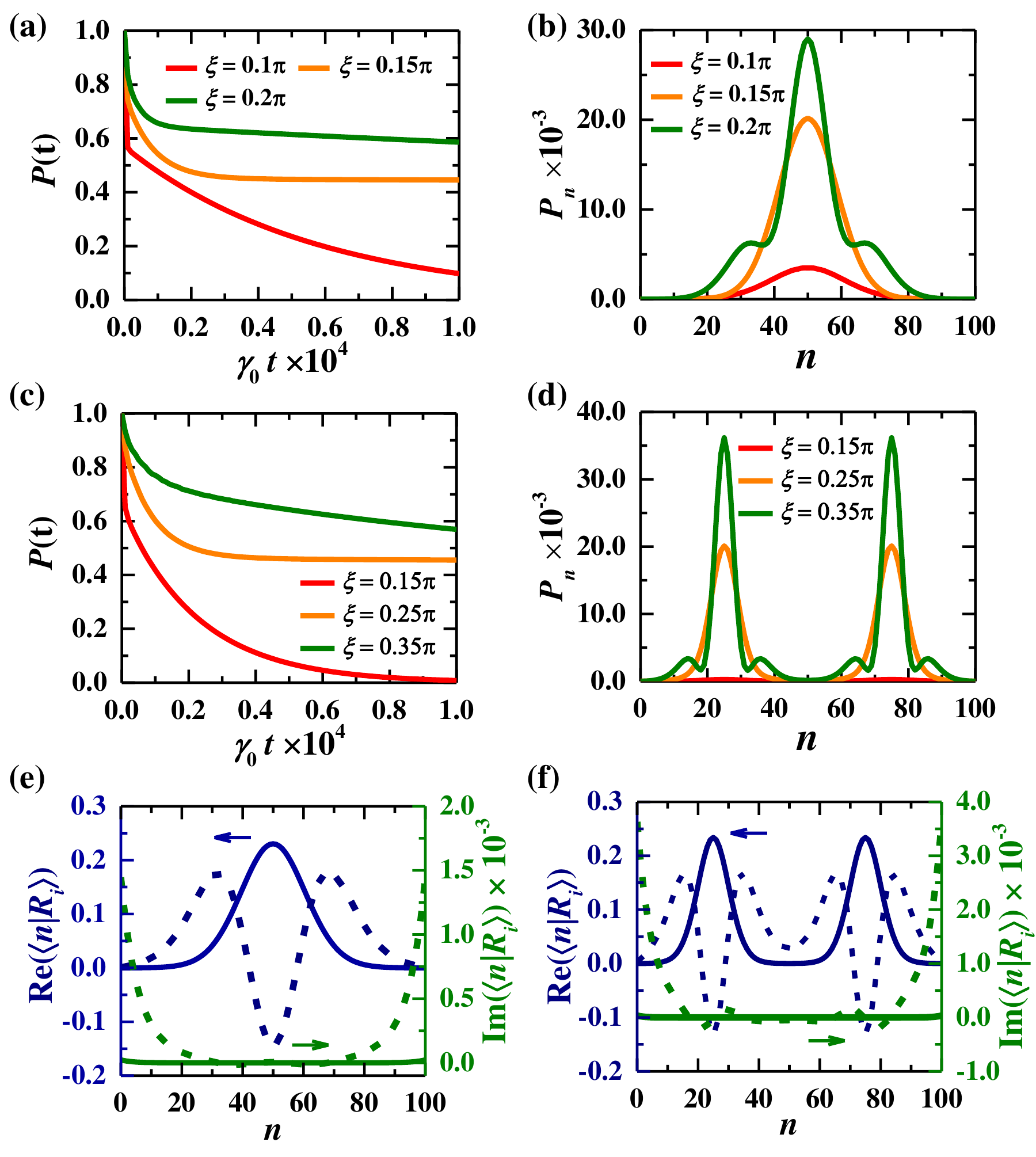}
\par\end{centering}
\caption{(a) Time evolution of the remaining population in a single-Gaussian
configuration for $\xi=0.1\pi,0.15\pi,$ and $0.2\pi.$ (b) Population
distribution at time $\gamma_{0}t=10^{4}$ corresponding to (a). (c)
Temporal behavior of the remaining population in a double-Gaussian
configuration for $\xi=0.15\pi,0.25\pi,$ and $0.35\pi.$ (d) Population
distribution at $\gamma_{0}t=10^{4}$ for cases in (c). (e) Eigenmodes
contributing to the dynamics at $\gamma_{0}t=10^{4}$ for the single-Gaussian
case with $\xi=0.2\pi;$ exactly two long-lived modes are involved.
Solid and dashed lines represent the first mode and the second mode,
respectively, with real and imaginary parts shown on the left and
right axes. (f) Same as (e), but for the double-Gaussian configuration
with $\xi=0.35\pi.$ \label{fig:eig_1}}
\end{figure}

For the single-Gaussian configuration, defined as
\begin{equation}
y_{\mu}=2\eta\exp\left(-\frac{\left(\mu/N-1/2\right)^{2}}{2\sigma^{2}}\right)-\eta,\label{eq:single-gaussian}
\end{equation}
the remaining population $P\left(t\right)=\sum_{\mu}\left|a_{\mu}\left(t\right)\right|^{2}$
exhibits monotonic decay for $\xi=0.1\pi,$ as shown in Fig.~\ref{fig:eig_1}(a).
At $\xi=0.15\pi,$ the decay slows considerably, and a finite fraction
of the excitation persists at long times. When $\xi$ is increased
to $0.2\pi,$ an even larger fraction survives; however, the spatial
distribution $P_{n}\left(t\right)=\left|a_{n}\left(t\right)\right|^{2}$
at $\gamma_{0}t=10^{4}$ deviates significantly from the original
Gaussian profile {[} Fig.~\ref{fig:eig_1}(b){]}.

A similar trend is observed in the double-Gaussian configuration,
described by 

\begin{align}
y_{\mu}= & 2\eta\exp\left(-\frac{\left(\mu/N-1/4\right)^{2}}{2\sigma^{2}}\right)\nonumber \\
 & +2\eta\exp\left(-\frac{\left(\mu/N-3/4\right)^{2}}{2\sigma^{2}}\right)-\eta.\label{eq:double_gaussian}
\end{align}
At $\xi=0.25\pi,$ the decay becomes markedly slower, and a non-negligible portion of the excitation remains at long times
{[}Fig.~\ref{fig:eig_1}(c){]}. For larger values of $\xi,$ the
spatial profile becomes increasingly distorted {[}Fig.~\ref{fig:eig_1}(d){]}.

To uncover the mechanism behind the deviations in localization patterns
at long times, we analyze the eigenmodes of the effective non-Hermitian
Hamiltonian $H_{\text{eff}}$ {[}Eq.~(\ref{eq:effective Hamiltonian}){]}.
We compute its complex eigenvalues $\lambda_{i}$ and corresponding
biorthogonal eigenvectors $\left\langle L_{i}\right|$ and $\left|R_{i}\right\rangle ,$
and project the initial state onto the eigenbasis as $c_{i}=\left\langle L_{i}\mid\psi\left(0\right)\right\rangle .$
The time-dependent contribution of each mode is given by $w_{i}\left(t\right)=\left|c_{i}e^{\lambda_{i}t}\right|^{2},$
allowing us to identify how many modes remain populated at late times.

In the single-Gaussian case, for $\xi=0.1\pi$ and $0.15\pi,$ only
one long-lived mode contributes significantly, with decay rates $\text{Im}\left(\lambda\right)=-8.8\times10^{-5}\gamma_{0}$
and $-4.9\times10^{-7}\gamma_{0},$ respectively. The corresponding
population contributions at $\gamma_{0}t=10^{4}$ are $w\left(t\right)=0.10$
and $0.45.$ At $\xi=0.2\pi,$ the dynamics involve exactly two modes,
with $\text{Im}\left(\lambda_{1}\right)=-3.3\times10^{-9}\gamma_{0}$
and $\text{Im}\left(\lambda_{2}\right)=-1.3\times10^{-5}\gamma_{0},$
leading to deformation of the excitation profile {[}Fig.~\ref{fig:eig_1}(e){]}.
The contribution of the first mode is $w_{1}\left(t\right)=0.38,$
and the second mode contributes $w_{2}\left(t\right)=0.20.$

In the double-Gaussian case, the system is again dominated by a single
eigenmode at $\xi=0.15\pi$ and $\xi=0.25\pi,$ with decay rates $\text{Im}\left(\lambda\right)=-2.2\times10^{-4}\gamma_{0}$
and $-6.9\times10^{-7}\gamma_{0},$ respectively. The population contribution
at $\gamma_{0}t=10^{4}$ is $w\left(t\right)=0.0079$ at $\xi=0.15\pi$
and $w\left(t\right)=0.46$ at $\xi=0.25\pi.$ At $\xi=0.35\pi,$
two long-lived modes contribute, with $\text{Im}\left(\lambda_{1}\right)=-3.2\times10^{-9}\gamma_{0}$
and $\text{Im}\left(\lambda_{2}\right)=-3.1\times10^{-5}\gamma_{0},$
resulting in a visibly deformed spatial profile {[}Fig.~\ref{fig:eig_1}(f){]}.
The contribution of the first mode is $w_{1}\left(t\right)=0.38,$
and the second mode is $w_{2}\left(t\right)=0.19.$

These results show that the geometry of the emitter array significantly
affects long-time localization by shaping the system's spectral structure.
Specifically, subradiant eigenmodes with exceptionally small decay
rates can remain populated over extended time scales, enabling robust
excitation trapping.

\subsection{Localization Control via Dominant Eigenmode under Transverse Modulation}

\begin{figure*}
\begin{centering}
\includegraphics[width=15cm]{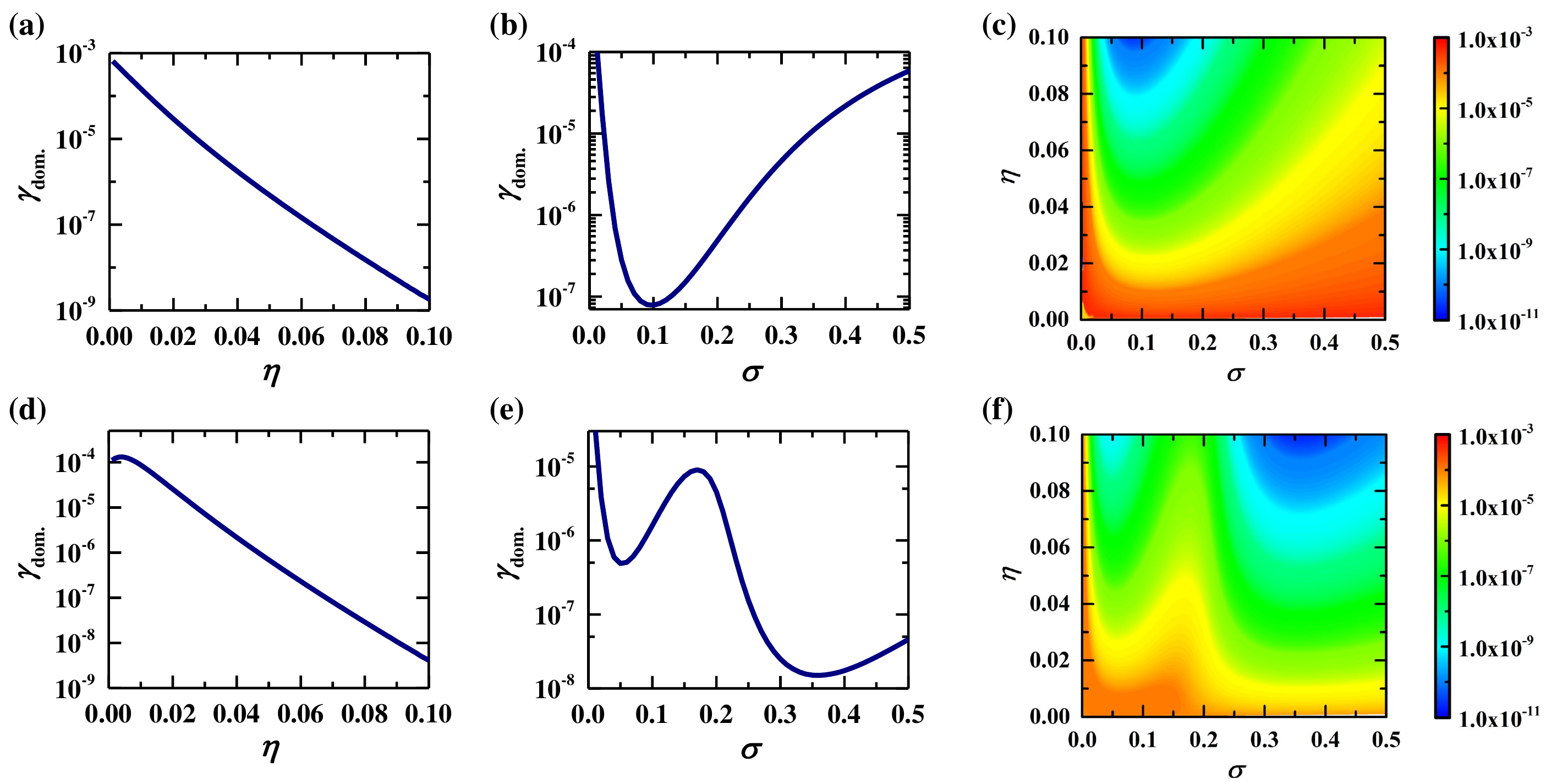}
\par\end{centering}
\caption{(a), (b) Decay rate of the dominant eigenmode of the non-Hermitian
Hamiltonian $H_{\text{eff}}$ for emitters arranged in a single-Gaussian
transverse profile {[}Eq. (\ref{eq:single-gaussian}){]}, with $\xi=0.15\pi.$
In (a), $\sigma=0.2$ is fixed while $\eta$ is varied, showing an
exponential suppression of the decay rate as $\eta$ increases. In
(b), $\eta=0.05$ is fixed and the decay rate exhibits a minimum around
$\sigma=0.1,$ suggesting an optimal configuration for minimizing
radiative loss. (c) Colormap of the decay rate as a function of both
$\sigma$ and $\eta.$ (d), (e) Decay rate of the dominant eigenmode
for emitters arranged in a double-Gaussian transverse profile {[}Eq.
(\ref{eq:double_gaussian}){]}, with $\xi=0.25\pi.$ In (d), $\sigma=0.075$
is fixed and $\eta$ is varied, leading to a monotonic reduction in
decay rate with increasing $\eta.$ In (e), $\eta=0.05$ is fixed
and the decay rate shows a local minimum near $\sigma=0.05.$ A secondary
minimum appears around $\sigma=0.35,$ configuration no longer retains
the intended double-Gaussian structure. (f) Colormap of the decay
rate across the $(\sigma,\eta)$ parameter space, identifying regions
that support strongly subradiant dominant modes. \label{fig:eig}}
\end{figure*}

To systematically explore how emitter geometry affects localization,
we focus on the dominant eigenmode\textemdash defined as the mode
with the largest contribution $\text{max}\left(w_{i}\left(t\right)\right)$
at $\gamma_{0}t=10^{4}.$ This mode is primarily responsible for the
residual excitation observed at long times.

We examine how the decay rate of this dominant eigenmode, $\gamma_{\text{dom.}}=\left.\text{Im}\left(\lambda_{i}\right)\right|_{\text{max}\left(w_{i}\left(t\right)\right)},$
depends on the transverse modulation parameters $\eta$ and $\sigma$
that define the single- and double-Gaussian profiles {[}Eqs.~(\ref{eq:single-gaussian})
and (\ref{eq:double_gaussian}){]}. These geometries control the spatial
variation of the emitter-waveguide coupling, which in turn influences
the formation of subradiant modes.

For the single-Gaussian case at fixed $\xi=0.15\pi,$ increasing $\eta$
while holding $\sigma=0.2$ leads to an exponential suppression of
the decay rate, as shown in Fig.~\ref{fig:eig}(a), indicating enhanced
localization due to reduced collective coupling. When $\eta=0.05$
is held fixed and $\sigma$ is varied, the decay rate reaches a minimum
near $\sigma=0.1,$ as seen in Fig.~\ref{fig:eig}(b). For larger
$\sigma,$ the smoother profile weakens localization as the geometry
approaches a uniform linear array. The full parameter dependence is
illustrated in Fig.~\ref{fig:eig}(c), where a distinct region of
minimized decay identifies an optimal regime for long-lived excitation
trapping.

A similar analysis for the double-Gaussian configuration at $\xi=0.25\pi$
yields analogous trends. Increasing $\eta$ at fixed $\sigma=0.075$
suppresses the decay rate monotonically {[}Fig.~\ref{fig:eig}(d){]}.
When $\eta=0.05$ is fixed and $\sigma$ is varied, the decay rate
exhibits a local minimum around $\sigma=0.05,$ as shown in Fig.~\ref{fig:eig}(e).
Although an even smaller decay rate appears around $\sigma=0.35,$
this configuration no longer preserves the double-Gaussian structure;
instead, the profile effectively becomes a single broad Gaussian,
losing the intended geometric separation. The full parameter dependence
is summarized in the colormap shown in Fig.~\ref{fig:eig}(f), which
maps the dominant decay rate across the $\left(\eta,\sigma\right)$
space. Strong localization is achieved within a narrow region where
the spatial configuration maintains well-separated peaks and sufficient
transverse displacement ($\sigma\sim0.05$).

These results demonstrate that by tuning the transverse geometry of
the emitter array, one can control both the lifetime and spatial structure
of the dominant eigenmode. This provides a powerful route to engineering
long-lived, spatially localized excitations in waveguide QED systems.

\section{Effect of Disorder on Different Emitter Arrangements}

In this section, we study how the introduction of disorder affects
the localization properties of emitter arrays that are originally
structured to support geometry-induced localization. We compare three
representative spatial configurations: a linear array, a single-Gaussian
profile, and a double-Gaussian profile. The system parameters are
chosen based on the findings from the previous section: for both the
linear and single-Gaussian arrangements, we consider $N=101$ emitters
with an inter-emitter spacing of $\xi=0.15\pi.$ This choice ensures
that, in the absence of disorder, the single-Gaussian configuration
supports localization dominated by a single long-lived eigenmode at
long times. By restricting the dynamics to a single mode, we avoid
complications arising from multi-mode interference, thereby facilitating
a clearer analysis of disorder effects. In contrast, for the double-Gaussian
arrangement, the same inter-emitter spacing yields negligible long-time
population; thus, we adopt $N=151$ emitters to enhance dipole-dipole
interactions and maintain sufficient population trapping. These settings
allow a consistent and meaningful comparison of disorder-induced modifications
across different geometries.

\subsection*{A. Linear Arrangement}

\begin{figure}[h]
\begin{centering}
\includegraphics[width=8.8cm]{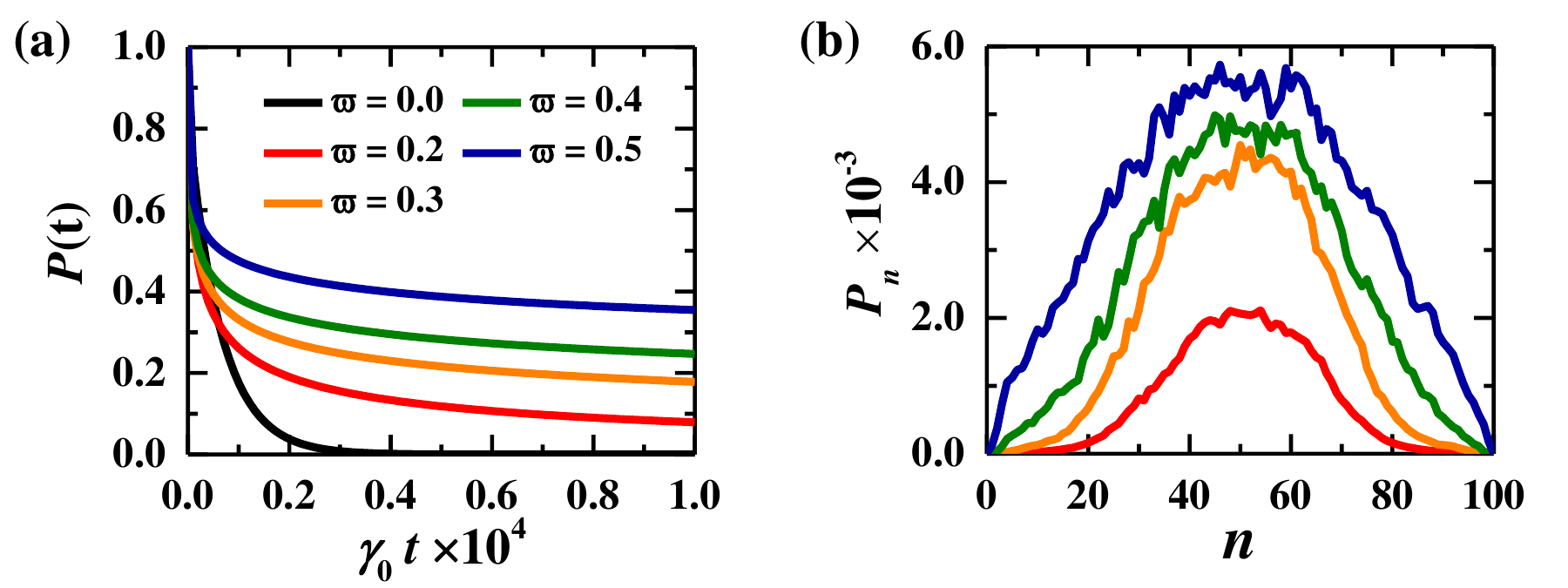}
\par\end{centering}
\caption{(a) Time evolution of the total remaining population in a linearly
arranged emitter array under increasing disorder. The system comprises
$N=101$ uniformly spaced emitters with inter-emitter spacing $\xi=0.15\pi$
and zero transverse displacement. (b) Spatial population distributions
at $\gamma_{0}t=10^{4}$ for selected disorder strengths. Parameter
values and color coding match those used in (a). \label{fig:linear_arrangement}}
\end{figure}

We begin by analyzing a linear array of two-level quantum emitters
initialized in a symmetric Dicke state. In this configuration, the
transverse displacement is set to zero for all emitters, such that
$y_{\mu}=0.$ The time evolution of the remaining population $P\left(t\right)$
under increasing disorder $W_{\mu}$ is shown in Figs.~\ref{fig:linear_arrangement}(a).
As the disorder strength increases, a larger fraction of the excitation
remains in the system, indicating a progressive suppression of radiative
decay.

The corresponding spatial distribution of the excitation at a long
time $\gamma_{0}t=10^{4}$ is shown in Fig.~\ref{fig:linear_arrangement}(b).
For weak disorder, the spatial profile remains approximately Gaussian,
while for stronger disorder, it becomes irregular but remains concentrated
near the center of the chain. These results suggest that disorder
leads to the formation of more localized and longer-lived subradiant
modes.

\subsection*{B. Single-Gaussian Arrangement}

\begin{figure}[h]
\begin{centering}
\includegraphics[width=8.8cm]{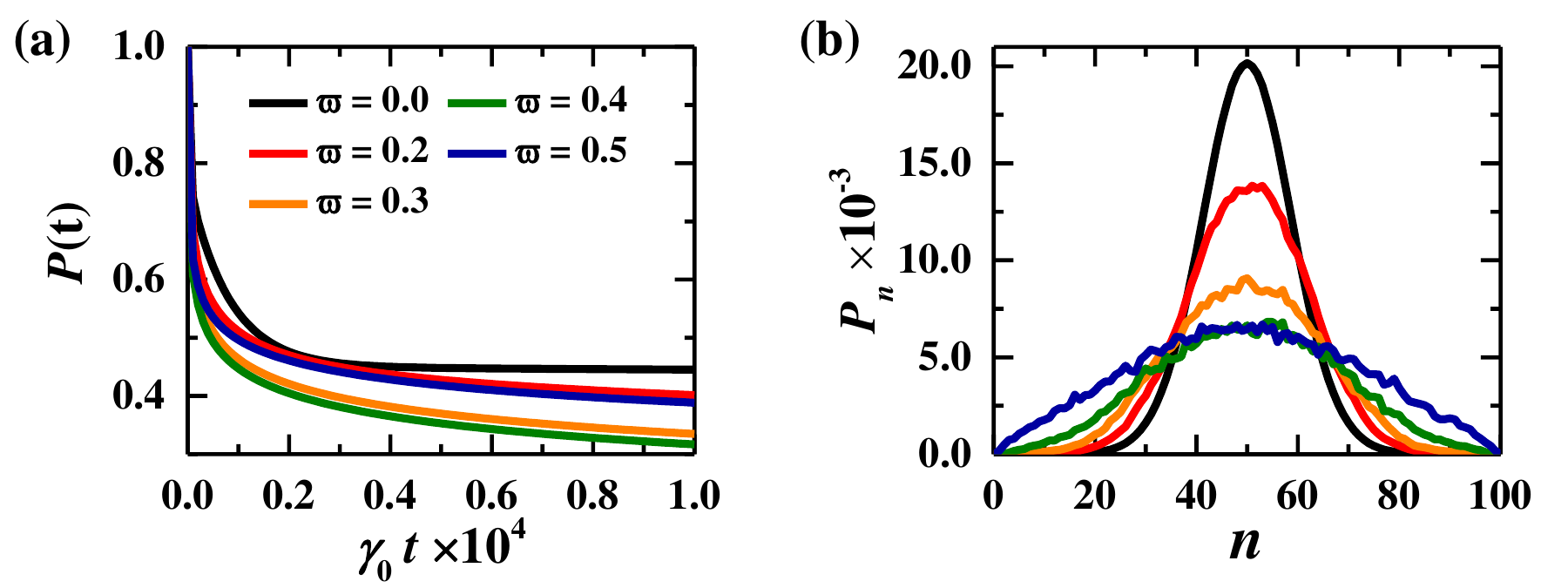}
\par\end{centering}
\caption{(a) Time evolution of the total remaining population for emitters
arranged in a single-Gaussian transverse profile, with parameters
$\eta=0.05,\sigma=0.2,N=101,$ and $\xi=0.15\pi.$ Moderate disorder
reduces the effect of geometry-induced localization, while stronger
disorder restores trapping via disorder-induced localization. (b)
Population distributions at $\gamma_{0}t=10^{4}$ for selected disorder
strengths. In the absence of disorder, spatial modulation of the coupling
leads to geometry-induced localization near the center of the array.
Parameter values and color coding match those used in (a). \label{fig:gaussian_arrangement}}
\end{figure}

We next consider a single-Gaussian emitter arrangement defined in
Eq.~(\ref{eq:single-gaussian}), with $\eta=0.05$ and $\sigma=0.2.$
In this configuration, the emitters remain uniformly spaced along
the propagation axis, while the transverse displacements create a
smoothly varying emitter-waveguide coupling strength across the array.
The time evolution of the remaining population $P\left(t\right)$
under increasing disorder $W_{\mu}$ is shown in Fig.~\ref{fig:gaussian_arrangement}(a).
Even in the absence of disorder, significant population trapping occurs
due to the site-dependent coupling strengths, which suppress collective
emission and lead to the formation of a long-lived localized mode.
As disorder is introduced, the trapped population initially decreases,
indicating that geometry-induced localization is sensitive to perturbations.
Upon further increasing the disorder strength, the population begins
to recover and eventually surpasses the clean-case level, suggesting
a crossover from geometry-induced to disorder-induced localization.

The spatial excitation distributions at a long time of $\gamma_{0}t=10^{4}$
provide additional insight into the effect of disorder. At zero disorder,
the population profile resembles a smooth Gaussian, reflecting the
underlying symmetry of the emitter geometry. As disorder strength
increases, the profile becomes increasingly irregular while remaining
centered around the middle of the array, as shown in Figs.~\ref{fig:gaussian_arrangement}(b).

While a crossover from geometry-induced to disorder-induced localization
is also present in the single-Gaussian configuration, this transition
is not readily visible in the spatial distributions. Both mechanisms
tend to produce centrally localized excitation profiles, making it
difficult to infer the dominant localization mechanism based solely
on the shape of the distribution. This limitation motivates the introduction
of a more structured geometry\textemdash namely, the double-Gaussian
profile\textemdash which enables a clearer visualization of the localization
crossover.

\subsection*{C. Double-Gaussian Arrangement}

\begin{figure}[h]
\begin{centering}
\includegraphics[width=8.8cm]{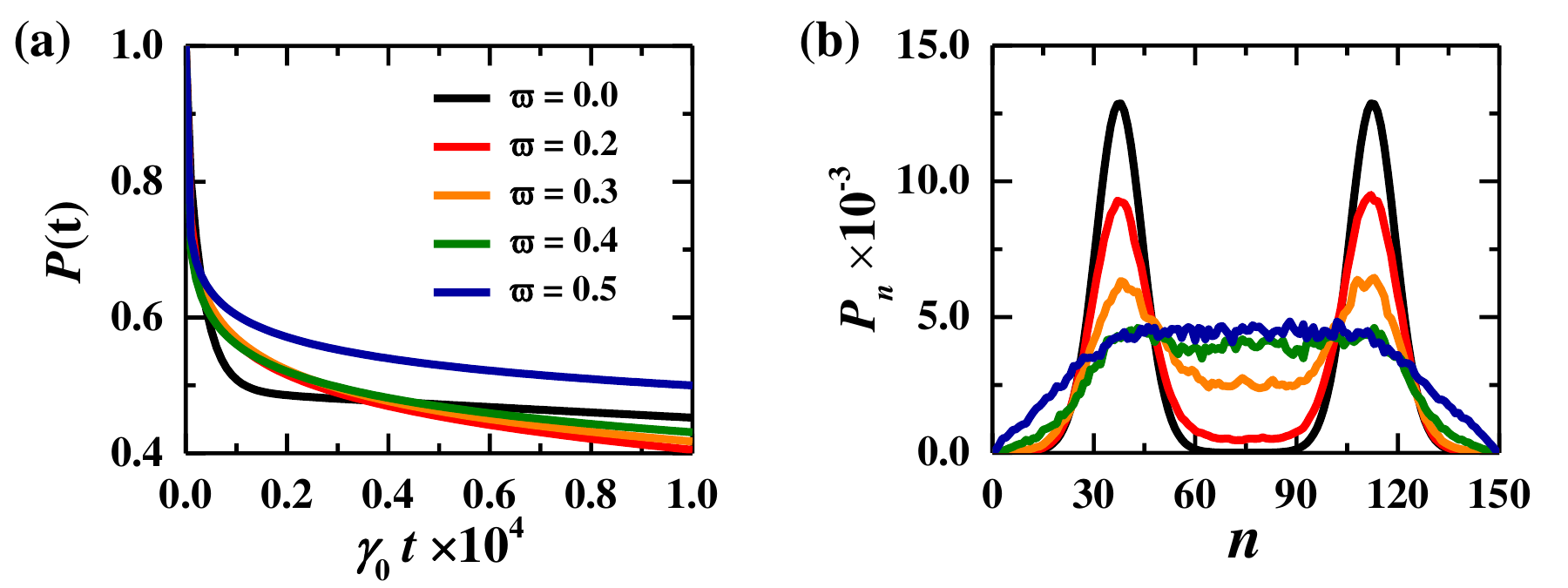}
\par\end{centering}
\caption{(a) Time evolution of the total remaining population for emitters
arranged in a double-Gaussian transverse profile, with parameters
$\eta=0.05,\sigma=0.075,N=151$ and $\xi=0.15\pi.$ Results are shown
under increasing axial disorder. (b) Spatial excitation distributions
at $\gamma_{0}t=10^{4}$ for selected disorder strengths. In the clean
limit, excitation is localized in two distinct regions corresponding
to weakly coupled emitter clusters. As disorder increases, the peaks
become less well-defined and partially overlap, indicating a deformation
of the original geometry-induced profile. \label{fig:double_gaussian}}
\end{figure}

We finally examine the double-Gaussian emitter arrangement defined
in Eq.~(\ref{eq:double_gaussian}), with $\eta=0.05$ and $\sigma=0.075.$
In this setup, the emitters are spatially configured to create two
regions of weak coupling to the waveguide, producing distinct localization
behavior. The evolution of the remaining population $P\left(t\right)$
under varying disorder strength is depicted in Figs.~\ref{fig:double_gaussian}(a).
Upon the introduction of disorder, the trapped population initially
diminishes, indicating that geometry-induced localization is sensitive
to perturbations. As the disorder strength increases further, the
population begins to recover and eventually exceeds the clean-case
value, consistent with a crossover from geometry-driven to disorder-driven
localization.

Further insights into the dynamics are revealed by examining the spatial
excitation profile at a long evolution time of $\gamma_{0}t=10^{4}.$
In the clean limit ($\bar{\omega}\approx0$), the distribution exhibits
two clearly resolved Gaussian-like peaks, corresponding to the engineered
regions of minimal coupling. As disorder increases, the initially
well-separated peaks become less distinct and gradually merge into
a single broad distribution centered in the array, as shown in Fig.~\ref{fig:double_gaussian}(c).
This transformation captures a transition between distinct localization
mechanisms: at low disorder, excitation confinement is governed by
geometry-induced subradiance, while at higher disorder, random scattering
dominates and drives disorder-induced localization.

\section{Size effects}

\begin{figure}[h]
\begin{centering}
\includegraphics[width=8.8cm]{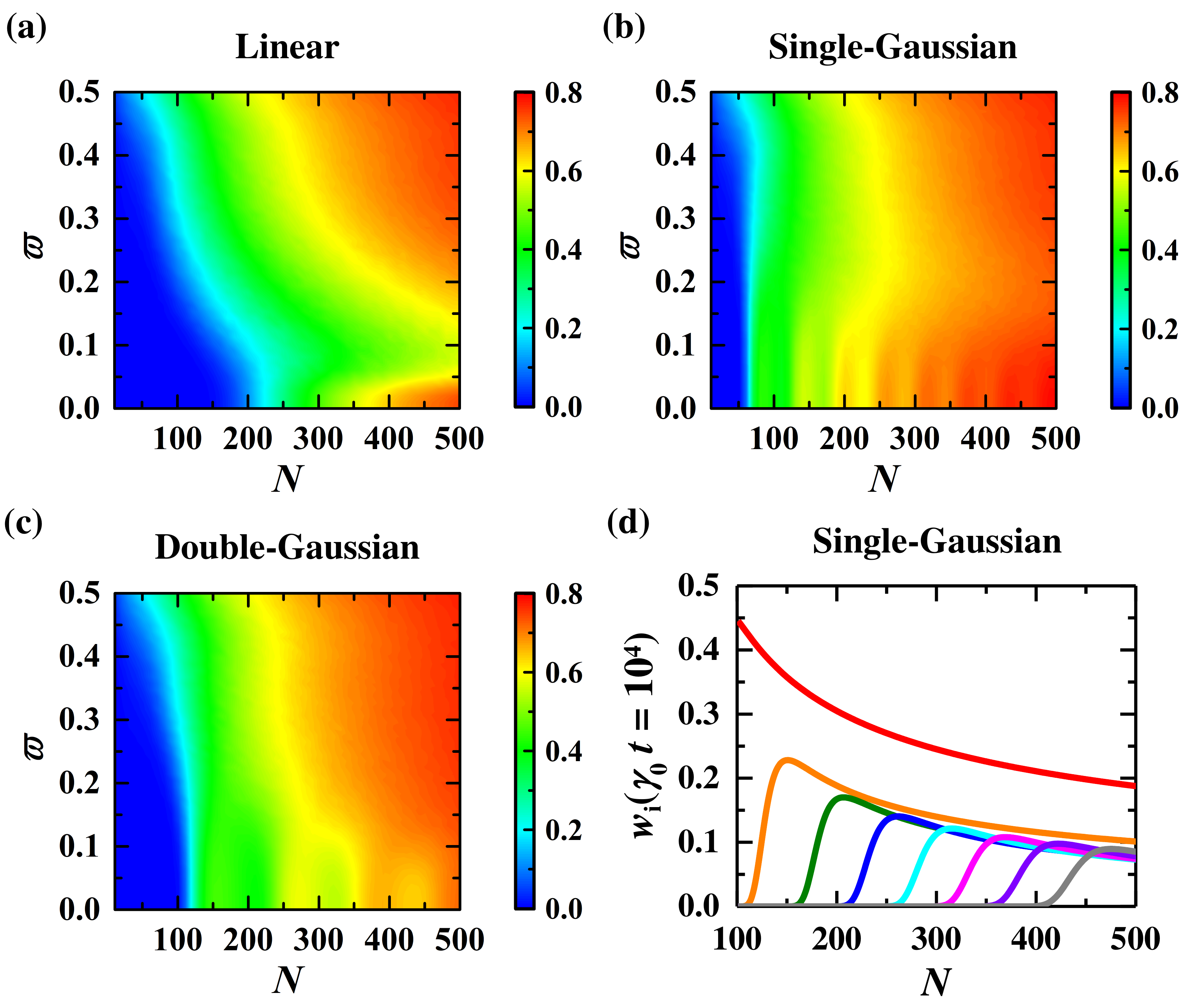}
\par\end{centering}
\caption{Total population remaining at time $\gamma_{0}t=10^{4}$ as a function
of emitter number $N$ and disorder strength $W_{\mu}$ for three
emitter configurations: (a) linear, (b) single-Gaussian, and (c) double-Gaussian.
(d) Contribution at $\gamma_{0}t=10^{4}$ as the emitter number $N$
changes. The red line represents the contribution of the 1st mode,
orange is for the 2nd mode, green for the 3rd mode, blue for the 4th
mode, cyan for the 5th mode, magenta for the 6th mode, purple for
the 7th mode, and gray for the 8th mode.\label{fig:size}}
\end{figure}

To investigate how system size influences population trapping in the
presence of disorder, we examine the total excitation population remaining
at a long time $\gamma_{0}t=10^{4}$ as a function of the number of
emitters $N.$ Results for three emitter configurations\textemdash linear,
single-Gaussian, and double-Gaussian\textemdash are shown in Fig.~\ref{fig:size}.
All three configurations are based on a fixed inter-emitter spacing
of $\xi=0.15\pi.$ The single-Gaussian profile is defined by transverse
modulation parameters $\eta=0.05$ and $\sigma=0.2,$ while the double-Gaussian
configuration uses $\eta=0.05$ and $\sigma=0.075.$

In the linear configuration {[}Figures~\textcolor{red}{\ref{fig:size}}(a){]},
population trapping is unexpectedly strong for large $N$ in the clean
limit due to subradiant modes supported by the spatial symmetry of
the system. These modes originate from destructive interference that
inhibits collective decay. However, even weak disorder breaks this
symmetry and rapidly suppresses the subradiant behavior. At higher
disorder strengths, Anderson-like localization becomes the dominant
mechanism, leading to a recovery in population trapping.

The single-Gaussian configuration {[}Figures~\textcolor{red}{\ref{fig:size}}(b){]}
exhibits more stable population retention across a wide range of system
sizes, particularly at low to moderate disorder strengths. This robustness
arises from geometry-induced spatial variation in emitter-waveguide
coupling. In this profile, emitters near the center are transversely
displaced furthest from the waveguide and therefore couple only weakly
to the photonic reservoir. As the system size increases, emitters
are added throughout the array, including both central and peripheral
regions. However, the long-time dynamics remain dominated by excitation
confined to the weakly coupled central region. Emitters located farther
from the center experience stronger coupling and tend to decay more
rapidly, contributing little to the population that persists at late
times.

The double-Gaussian configuration {[}Figures~\textcolor{red}{\ref{fig:size}}(c){]}
exhibits population trapping that depends sensitively on system size
and disorder strength. At low disorder strengths, the system effectively
behaves as two spatially separated single-Gaussian segments, and therefore
requires a larger number of emitters for significant population trapping.
As disorder increases, the distinction between the two lobes becomes
less significant, and the system begins to support long-lived modes
with enhanced excitation confinement. This transition reflects a shift
in the dominant localization mechanism, from geometry-induced subradiance
to disorder-induced localization.

In both the single-Gaussian and double-Gaussian configurations, ripples
in the remaining population are observed in the clean limit as the
number of emitters $N$ increases {[}Figures~\textcolor{red}{\ref{fig:size}}(b)
and (c){]}. This behavior can be attributed to two factors: first,
the increasing number of subradiant modes as $N$ grows, which leads
to a redistribution of the population among these modes; second, some
of these subradiant modes experience a decrease in their decay rates.
These two factors combined cause the trapped population to exhibit
ripples as the system size increases. For the single-Gaussian configuration,
these ripples are most pronounced when $N$ is in the range of $100$
to $350,$ where the number of remaining long-lived modes at $\gamma_{0}t=10^{4}$
increases from one to six {[}Figures~\textcolor{red}{\ref{fig:size}}(d){]}.
As more long-lived modes become available, the population becomes
more evenly distributed among them, leading to a reduction in the
amplitude of the ripples. This explains why ripples are no longer
observed for $N\gtrsim350,$ where a sufficiently large number of
modes results in a smooth population distribution. Additionally, as
$N$ increases, the number of subradiant modes contributing to the
dynamics also increases, and the position of the ripple aligns with
the peak value of the $i\text{th}$ subradiant mode's contribution
{[}Figures~\textcolor{red}{\ref{fig:size}}(b) and (d){]}, showing
that the excitation trapping is directly related to the contribution
of these modes. A similar trend is observed in the double-Gaussian
configuration. However, as disorder increases, this effect diminishes.
Since disorder lacks a specific geometric structure, the population
distribution becomes more uniform on average, and the ripple effect
vanishes as the system becomes dominated by disorder-induced localization.

\section{Discussion and conclusion}

In this work, we have investigated excitation localization and population
trapping in one-dimensional arrays of quantum emitters coupled to
a waveguide, focusing the interplay between engineered emitter geometries
and static disorder. Using time-domain simulations combined with biorthogonal
eigenspectrum analysis of the non-Hermitian effective Hamiltonian
\citep{Giusteri2015}, we identified two distinct mechanisms of localization
in open quantum systems: geometry-induced subradiance and disorder-induced
Anderson-like localization \citep{Jen2020,Jen2021,Jen2022,Wu2024}.

Our results show that spatially modulated emitter arrangements\textemdash such
as single- and double-Gaussian transverse profiles\textemdash can
support long-lived collective modes with significantly reduced decay
rates compared to uniform arrays. These subradiant modes arise from
inhomogeneous coupling to the waveguide and are highly sensitive to
the transverse emitter distribution. Through systematic parameter
mapping, we identified geometric regimes that optimize population
trapping and coherence retention.

In contrast, localization in uniform linear arrays requires sufficient
disorder to break spatial symmetry and enable interference-based confinement.
By varying disorder strength, we observed a crossover from geometry-induced
to disorder-induced localization. This transition is subtle in single-Gaussian
configurations, but becomes pronounced in double-Gaussian arrays,
where the spatial excitation pattern evolves from a double-peaked
profile to a broad, centrally localized distribution, indicating a
transition from geometry-induced to disorder-induced confinement.

We further examined the impact of system size, showing that geometry-induced
localization remains effective in large arrays. This suggests a practical
advantage over disorder-based or symmetry-protected localization schemes,
particularly in the context of scalability and experimental implementation.

These findings provide new insights into how spatial geometry and
disorder can be used as complementary tools to control decoherence
and dissipation in waveguide QED systems. The predicted localization
behavior and enhanced excitation lifetimes are directly relevant to
ongoing experiments with cold atoms coupled to nanophotonic structures
\citep{Kim2019,Luan2020}, superconducting qubits coupled via microwave
waveguides \citep{Roushan2017,Xu2018}, and solid-state emitters embedded
in photonic crystals \citep{Arcari2014}.

Looking ahead, several promising directions remain, including extending
the analysis to the multi-excitation regime and incorporating temporal
disorder or nonlinear interactions. In addition, introducing chiral
coupling\textemdash asymmetric emission into left- and right-propagating
modes\textemdash would provide a promising route to explore nonreciprocal
localization dynamics and direction-dependent subradiance. By bridging
theoretical predictions with practical implementations, our findings
pave the way for designing scalable, coherent quantum networks and
quantum memories with enhanced control over dissipation and decoherence.

\section*{Acknowledgments}

We acknowledge support from the National Science and Technology Council
(NSTC), Taiwan, under the Grant No. NSTC-112-2119-M-001-007, No. NSTC-112-2112-M-001-079-MY3, No. NSTC-113-2112-M-002-025, and No. NSTC-112-2112-M-002-001, and from Academia Sinica under Grant
AS-CDA-113-M04. We are also grateful for support from TG 1.2 of NCTS
at NTU.

\bibliographystyle{apsrev4-2}
\bibliography{localization}

\end{document}